# Simulating the conflict between reputation and profitability for online rating portals

Boris Galitsky and Mark Levene

School of Computer Science and Information Systems
Birkbeck College, University of London
Malet Street, London WC1E 7HX, UK
{b.galitsky, m.levene}@dcs.bbk.ac.uk

## Abstract

We simulate the process of possible interactions between a set of competitive services and a set of portals that provide online rating for these services. We argue that to have a profitable business, these portals are forced to have subscribed services that are rated by the portals. To satisfy the subscribing services, we make the assumption that the portals improve the rating of a given service by one unit per transaction that involves payment.

In this study we follow the "what-if" methodology, analysing strategies that a service may choose from to select the best portal for it to subscribe to, and strategies for a portal to accept the subscription such that its reputation loss, in terms of the integrity of its ratings, is minimised. We observe that the behaviour of the simulated agents in accordance to our model is quite natural from the real-would perspective. One conclusion from the simulations is that under reasonable conditions, if most of the services and rating portals in a given industry do not accept a subscription policy similar to the one indicated above, they will lose, respectively, their ratings and reputations, and, moreover the rating portals will have problems in making a profit. Our prediction is that the modern portal-rating based economy sector will eventually evolve into a subscription process similar to the one we suggest in this study, as an alternative to a business model based purely on advertising.

## Introduction

Portals providing online rating of services, such as financial services, are becoming more popular nowadays. A rating portal providing comparisons between competitive services, has the potential of becoming a well established web enterprise. For some services the comparison is performed based on a set of measurable values such as performance and price, for example when the service involves computer hardware. In such an environment, services can make a rational decision whether they wish to advertise on the portal, based on the set of measurable values. However, for some services like banking, brokerage and other financial services, characterised by such parameters as customer support quality, it is impossible to establish an objective set of measurable values. In these cases the rating portals publish their scores for the competing businesses based on their own private estimation strategy. We believe that evolution of the interactions between the agents being rated and rating agents is an important social process which is worth thorough simulation and understanding.

There are two common ways to rate services: (1) assigning to each service a score in a range of values, or (2) ranking the services in order of preference. In this study we simulate the plausible interaction between portals and services using a simplified model, and we analyse possible scenarios of how services can influence the portals' rating system. Our model is based on a straightforward revenue model for rating portals, where they require the rated services to be paying subscribers in order to obtain a rating. Within this model we follow the dynamics of how the competing services may influence the portal to improve their respective ratings.

This work follows along the lines of the study of an economy of web links, where the potential monetary values of web links has been explored and a link exchange process has been simulated (Galitsky and Levene 2003). Clearly, assuming that the majority of links are established as a result of such exchange is unrealistic, however, it sheds some light on how web links might be established in a future economy should the process of link exchange become prevalent. Analogously, in the current study, we overstate the role of the interaction between a service and a rating portal in order to judge how the former may affect the latter in the course of a competition for a better rating.

Web-based rating portals are normally assumed to be independent, and they do their upmost to impress on the customers of the services being rated that this is indeed the case. However, the current web economy does not broadly support the revenue model, where rating portals charge their customers, so instead, the companies competing for ratings fund the portals through advertising their products on the portal's web site. This model of advertising can be tied in to the rating mechanism leading to a new form of advertising. In some sense this is similar to the paid placement model of advertising on search engines, where the (sponsored) ranking for a given query is decided by a bidding process for keywords that will be submitted as queries by users to the search engine.

In this study we suggest a plausible model of transferring resources from services to rating portals in the form of a *rating subscription*. In this model services enter into a contract where they are pay rating portals for a *small increase* of their rating on the portal's list. In the situation where the subscription rate is the same for all participating services, the rating results would not significantly deviate from "pure" ones which are not sponsored by any services. Therefore, in this case, we can argue that the reputation of the rating portals would not be strongly affected by sponsoring services on their rating. The drawback of this scheme is that that the services, which decide not to join into a subscription agreement, will either be forced to be withdraw from the rating scheme or suffer low ratings.

The methodology of this study is as follows. We analyze the current business model of web portals that provide rating services and hypothesise that that would be willing to be funded by the services rather than by their customers. We then conduct the *what-if* study suggesting a simple model with rational agents for services and portals as possible for a simulation of the subscription model. The resultant behaviour is verified and analysed with respect to the possibility of extracting patterns of rating subscription-based behaviour from real publicly available data. We conclude the paper with a discussion of how the predicted subscription process fits into the current advertising models; also the process itself is considered from the standpoint of conflict resolution in multi-agent systems.

## Economic model

Portals are primarily characterised by their reputation. To express this quantitatively, we refer to the difference between the average rating of each service and the individual rating of each service on each portal. The higher the portal's reputation, the more potential customers it has and a higher the number of web surfers who would follow the portal's recommendation to select a particular (top-rated) service. Also, the higher the portal's reputation is, the higher is its appeal for the services to be rated by this portal, and, therefore, the potential revenue stream for the portal is higher. At the same time, when a portal accepts resources from the services it rates, its reputation may drop because its rating may become less objective. The dynamics of such a process is the subject of the current study.

How should the reputation of a portal be defined? Here we suggest a simple model where there is no *objective* rating: each portal, while having its own rating system, aims to *maximise* its revenues on the one hand, and on the other hand aims to deviate as little as possible from the *average* portal rating. The justification for this is that often the public perceives the average (or typical) rating (or opinion) as the most trustworthy. This is in contract to a distinctive or radical opinion, which may be too risky to follow.

We select our model for the average rating as being the "best" based on psychological studies of how the public perceive the parameters of relating to the subject of interest (see e.g. Myung and Pitt 2003). We also conducted the limited study of how financial services (mutual funds) are ranked according to search engine keyword-based queries and compared this data with the most popular rating of mutual funds according to morningstar.com. We observed that averaging is the simplest way to perceive the rating data, and that the most popular search engine (google.com) is quite close (however, not the closest) to the rating, averaged over the four search engines[ML4] (see the formal model section below and the comparison of web search ratings).

Our model reproduces the real-life conflict between the services and portals: each service is determined to improve its ratings irrespectively of how it affects a portal's reputation, and vice versa, each portal wishes to achieve higher reputation and at the same time to increases its revenues. No evident compromise is possible.

Rather than attempt to build an optimal strategy for services and portals, we suggest a simple rational strategy, where the agents only take into account two parameters, one concerning themselves and the other concerning the opponent agents. One way of verifying this approach is by changing the rational strategy into a random one and showing that, in this case, we obtain results which do not follow real-world phenomena. For example, if the strategy is random on either the services or portals side, then we observe the unnatural and irrational properties of the simulated behaviour. In the last part of the simulation section below, we will enumerate these properties.

To provide realistic initial conditions for our simulation, we have chosen thirteen mutual funds as financial services and four well-known portals, which provide ratings for these services. We have chosen a relatively small number of agents so that we can track their behaviour and observe the results. We have also verified the model with a larger set of participating agents and although the convergence time is longer, we recorded the same phenomenology as revealed by the smaller data set.

We have simulated all phases of the subscription process, including the initial phase, when the services initiate the subscription process to modify their initial rating, and the terminal phase, when the services run out of resources or see no further benefit in participating in the process. We believe that simulating the full cycle of the subscription process rather than just recording the resulting stationary process, provides us with sufficient phenomenology to identify similar processes in the real business world.

## A formal model

We use matrix $M$ to express ratings, where $M(s,p)$ denotes the rating of service $s$ by portal $p$. Ratings of services are represented by integers from 1 to $ns$, where the ratings are presented in ascending order from the highest rated service (1) to the lowest one ($ns$). Each column of $M$ contains integers $1,\ldots,ns$ in a certain order such that each integer occurs only once, i.e. a portal cannot assign the same rating to two services.

The average rating for a service, $s$, over the set of portals, is given by:

$$r_{avg}(s) = \sum_p \frac{M(s,p)}{\# p}$$

where $\#p$ denotes the number of portals. The reputation for a portal is calculated as the reciprocal of the deviation of the rating it gives to each service from the average rating of the service, and is given by

$$reput(p) = \frac{1}{\sum_s \left( M(s,p) - r_{avg}(s) \right)}$$

When choosing which portal to subscribe to a service chooses the portal with the highest reputation while taking into account its possible increase in rating so that its rating will be as close to the highest rating (i.e. 1) as possible. More specifically, service, $s$, makes a subscription offer to portal, p, such that

$$\frac{reput(p)}{M(s,p)}$$

is maximized.

Out of the totality of services which make a subscription offer to a given portal, the portal selects the one which would decrease its reputation the least. More specifically, portal, $p$, chooses to accept the subscription from the service, $s$, that minimizes

$$| M(s,p) - r_{avg}(s) |.$$

When portal, *p*, accepts the subscription offer from service, *s*, then *s* transfers *m* resource units to *p*, and *p* increases the ranking of *s* by one. So, if *s* was ranked at position *n* and *s'* was ranked at position *n-1*, their rankings are swapped. In the special case when *s* was already ranked at position 1, then the portal does not accept the offer from *s*.

For example, the top scenario shown in Figure 1 is beneficial for a rating portal because after the rating for the subscribed service is increased, this services rating will get closer to the average rating over all portals, and therefore its reputation increases as a result of the transaction. Conversely, for the scenario shown on the bottom of Figure 1, the increase in rating desired by this service will cause the portal's reputation to decrease, since its rating of the service moves further away from the average rating for this service.

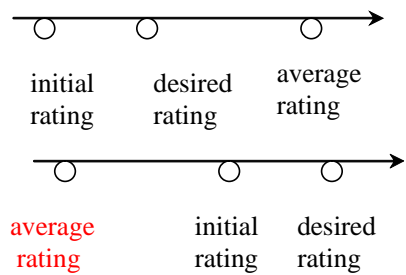

Figure 1: Two scenarios showing how portals' ratings change

The algorithmic steps of the simulation are depicted at Figure 2. Two modules where the selection strategies are implemented are highlighted by surrounding dotted lines.

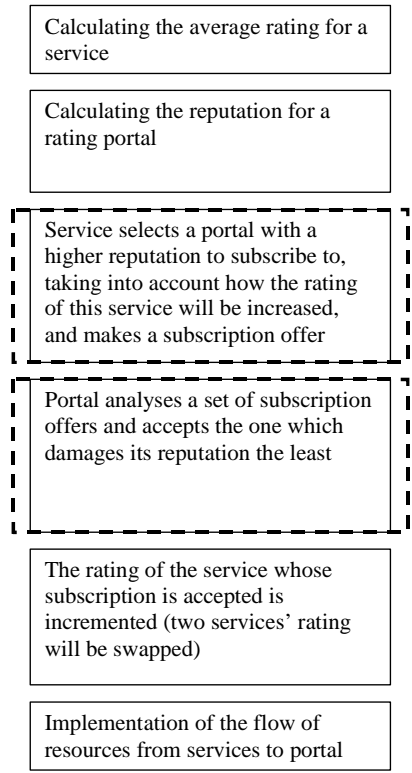

Figure 2: The modules of the interaction process.

The simulation that produced the results described in the next section was implemented in Matlab and is available from the first author on request[ML5].

## Simulation runs

We first present the dataset that we used to track the dynamics of the interaction between services and portals, capturing the behaviour patterns of the involved agents and judging their rationality. The purpose of this dataset is to verify the consistency of our model with respect to the rationality of the portal selection strategy of services and the offer acceptance strategy of portals.

We formed the initial dataset of ratings from a selected set of thirteen mutual funds, rated by a set of four portals as a 4 by 13 matrix, where each column, representing a portal, contains numbers from 1 to 13 (without repetitions) denoting the ratings of the services by the portal. To visualize the dynamics of the interactions, we plot the two following types of curves:

1) Distributions of ratings/reputations and resources of services and portals; and
2) The evolution of these parameters over time.

The first type of curve is useful to show how services and portals are different at a specific point in time, and the second type shows the changes of ratings/reputations and resources for each agent over a period of time.

In addition to the initial ratings, the following simulation parameters were used:
 1) Initial resources set at 1000 units.
 2) Subscription fee (per transaction) set at a flat rate of 50 units.

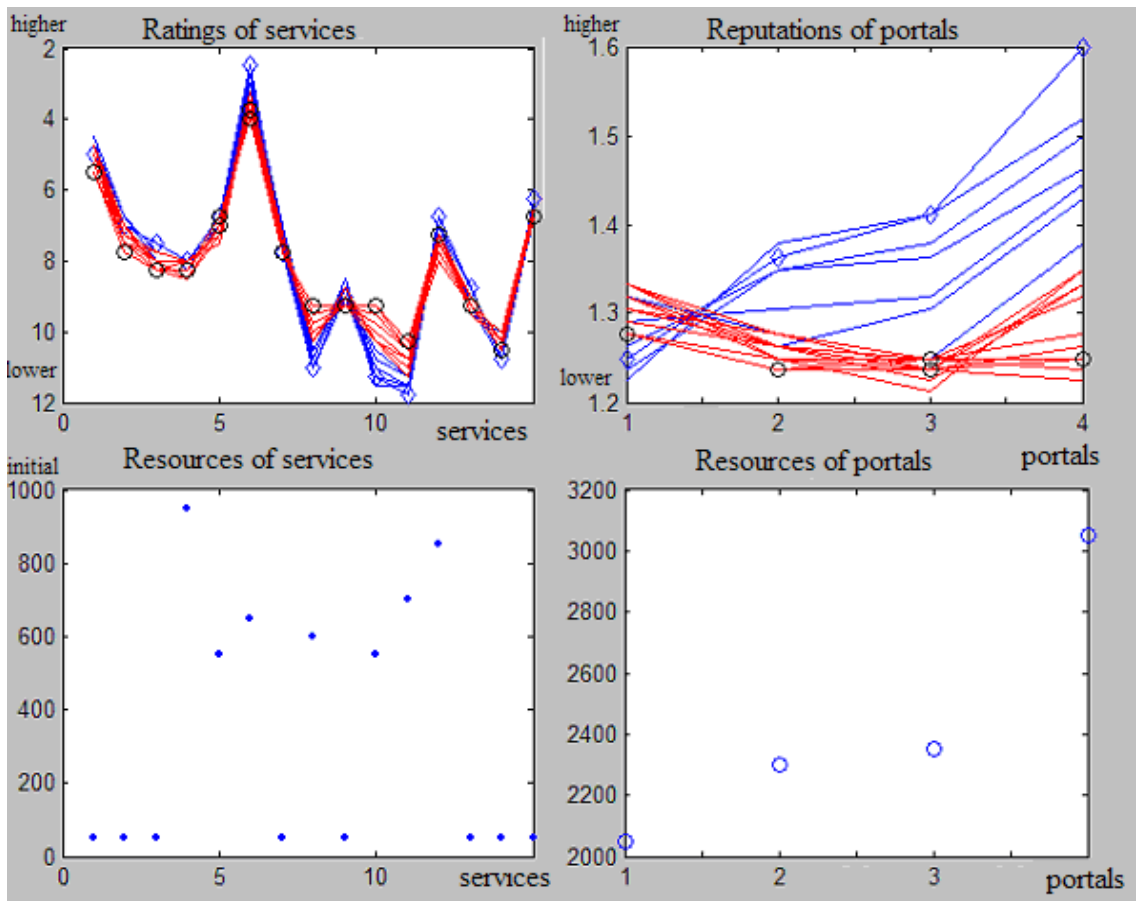

Figure3: Distributions of ratings and resources. Both rational strategies of service subscription offers and portals acceptance are used.

In  Figure 3, on the charts for  distributions of  reputations/reputations shown in the upper half of the figure,  diamonds denote the initial ratings and circles denote the final ones. On the charts for resources shown in the lower half of the figure, dots and circles denote the respective values for services (left) and for portals (right).

We observe that about a half of services have run out of resources all of which have been transferred over time to the portals (see the dots on the bottom of the resources of services chart). Remarkably, these services (except for #13) did not improve their ratings.

Those services which dramatically improved their ratings still have some resources left, which they can use to further improve their rating on portals. We observe on the chart that only those services whose average rating is below 11 have actually improved their rating.

Therefore, one may suppose that only the lowest rated services will have an interest in paying a subscription fee to the portals (assuming that an objective and independent rating is possible). However, this is far from the truth: the other services need to keep trying to move the ratings in the direction which favours them, otherwise, their rating will significantly deteriorate relative to their initial rating. What our simulations show is that the group of services with an initial higher ratings run out of resources earlier than the group with an initial lower rating. This happens, since the first group has to compensate for the actions of the second group.

Naturally, the sum of the average ratings of the services is constant irrespectively of individual ratings. However, this is not the case for portals, whose reputations get worse in the course of subscription process.

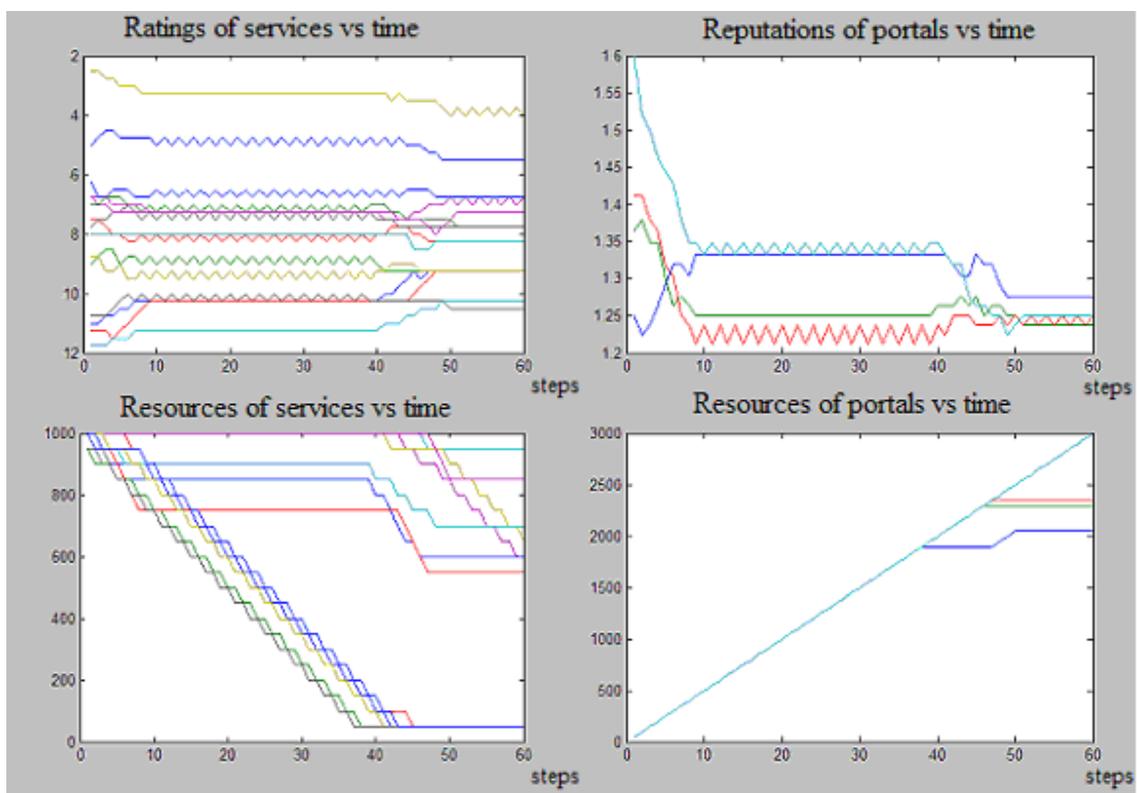

Figure 4: The evolution of ratings/reputations and resources of services and portals over time.

It takes first 10 steps to establish an equilibrium of ratings between the services and an equilibrium of reputations between the portals (see Figure 4). Once the equilibrium is achieved, an oscillation pattern appears, which is caused by pairs of financial services that have their ratings swapped between position $i$ and position $i$-1. As a result, the reputations of the portals are interchanged in a similar way, leading to an oscillating pattern between portals as well. The amplitude of oscillations for services is a quarter of unit (one out of four changes to the reputations of portals contributes to this amplitude).

On the other hand, for the portals we observe oscillations with amplitudes which are higher than a single unit.

There is the critical point, at steps 38-45, when the interaction between the agents changes, at the time when eight of the services run out of resources. After that, the offers of the remaining services are always accepted, and the portal reputations are subject to further deterioration, as well as the ratings of these eight services that ran out of resources. However, the ratings of those services which have not run out of resources during these steps increase during steps 45-60. After that time, there is a smaller number of services capable of paying a subscription fee; 3 out of 4 of the portals are not offered a subscription and therefore do not increase their resources after this critical point. The competition for the subscription offers by services to be accepted by portals is still strong: all services wish to subscribe to the same portal, and the portal they all desire to subscribe to can only accept the subscription from a single service according to the rules of the game.

We outline the five zones we have detected within the evolution charts of interacting services and portals:

1) The *equilibrium establishing zone*;
2) The *oscillation zone*;
3) The *resources disappearance zone*;
4) The *limited resources equilibrium establishing zone*; and
5) The *stationary zone*.

As is visible in the evolution charts (Figure 3), in accordance to what we have revealed in the distribution charts (Figure 4), only the lowest-rated services benefit from the process (the bottom part of top-left chart in Figure 3). The evolution charts show that the rating of the lowest-rated services increases during both the first (oscillation) and the fourth (limited resources equilibrium establishing) zones. There is no service that would significantly benefit from the process, since no service has improved its average rating by more than 2 units.

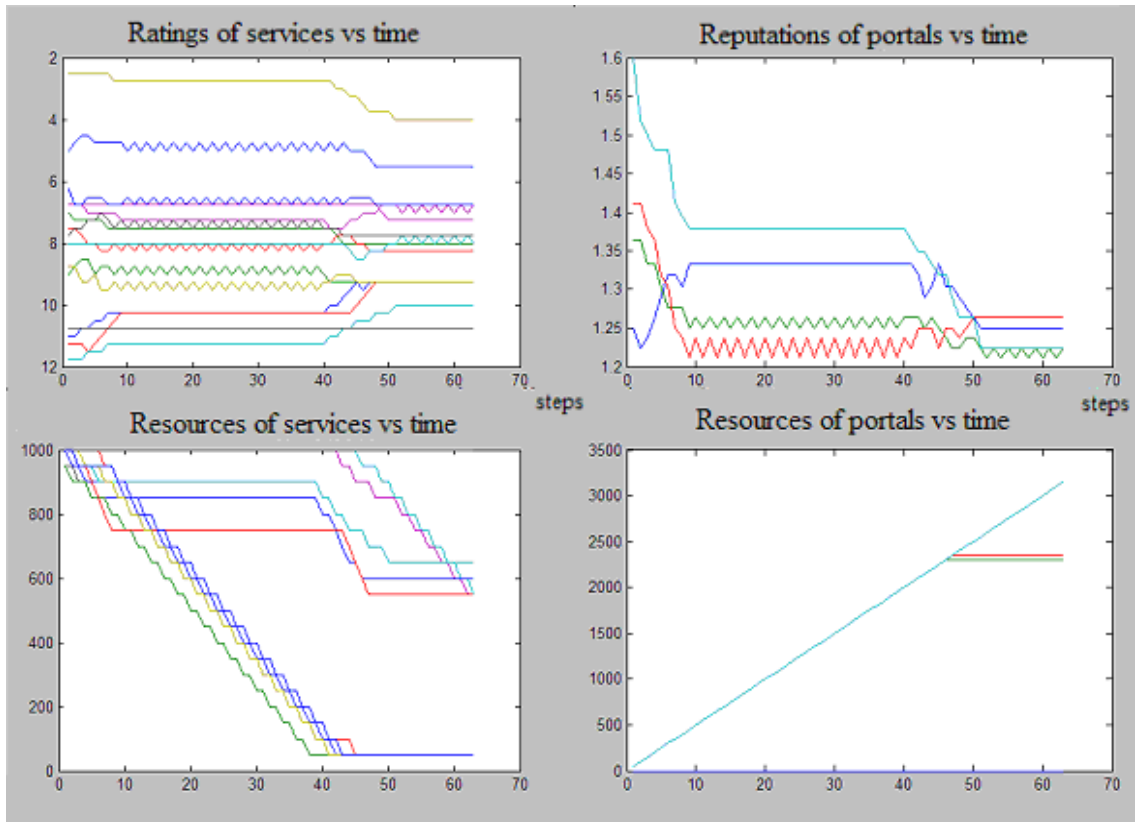

Figure 5: The evolution of ratings/reputations and resources of services and portals over time, where one portal with a low initial reputation is independent (i.e. it does not accept service subscription).

When a given portal does not accept subscription fees, its rating in the evolution curve in an environment where other portals accepts subscription fees is quite similar to the situation above, where every portal accepts subscription fees. The resource curve for this portal is a horizontal line on the bottom of the chart; the three remaining resources curves go together until step 48 when two of the portals stop gaining any further resources.

The resultant reputation of a portal is even lower when no subscription can be accepted, because the objective ratings it publishes will have a stronger deviation from the average, which is mostly affected by the portals that can accept subscriptions. The reputation dynamics closely follow the case when this portal can accept a subscription (see Figure 4). Therefore, the overall subscription process is only weakly affected by a minority of portals which cannot accept subscription.

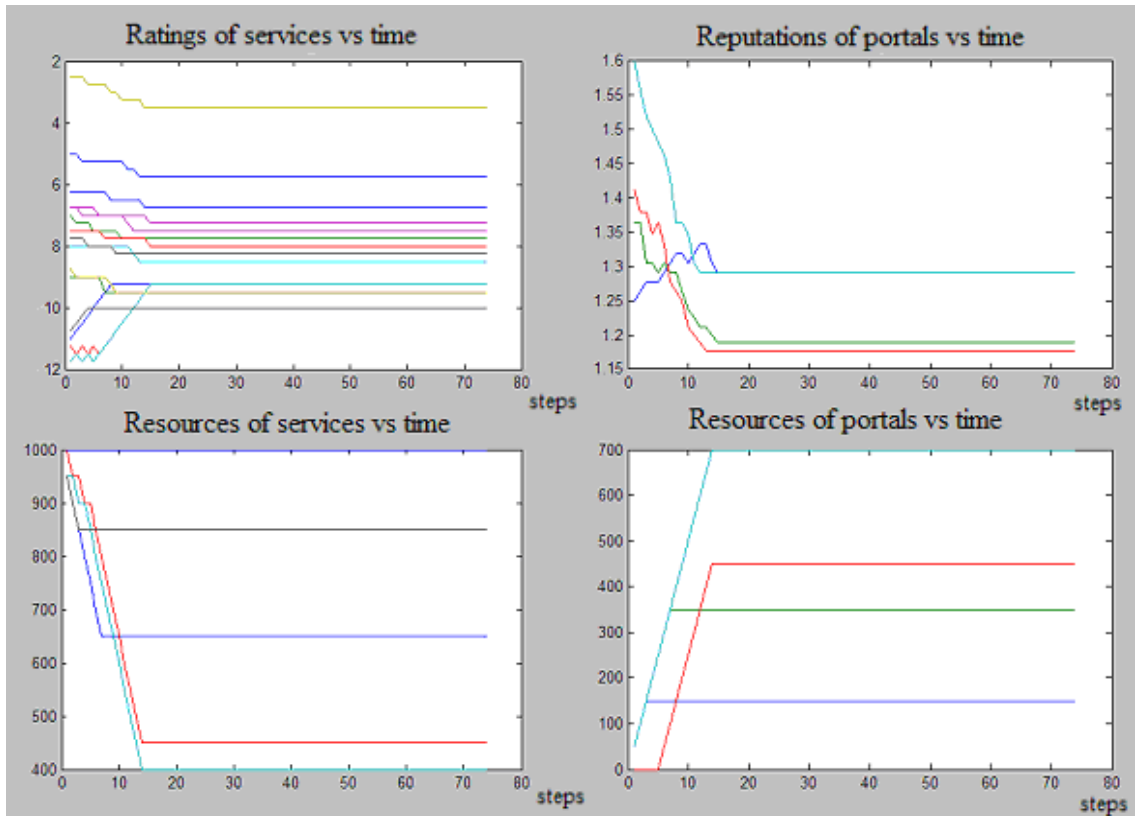

Figure. 6: The evolution curves where only the lowest-rated service subscribes

When only the lowest-rated sites choose to offer subscription fees to portals to increase their ratings, the process immediately converges to the fifth zone (stationary) without passing through the intermediate zones (see Figure 6). In this case portal reputations significantly drop, as well as the ratings of all the services which decided to avoid the subscription. The case when some service withdrew from the subscription process is quite different from that of portals: there is a dramatic change in the process for the former, whereas the latter case does not significantly change. The competition between the services is not strong enough to lead to an oscillation.

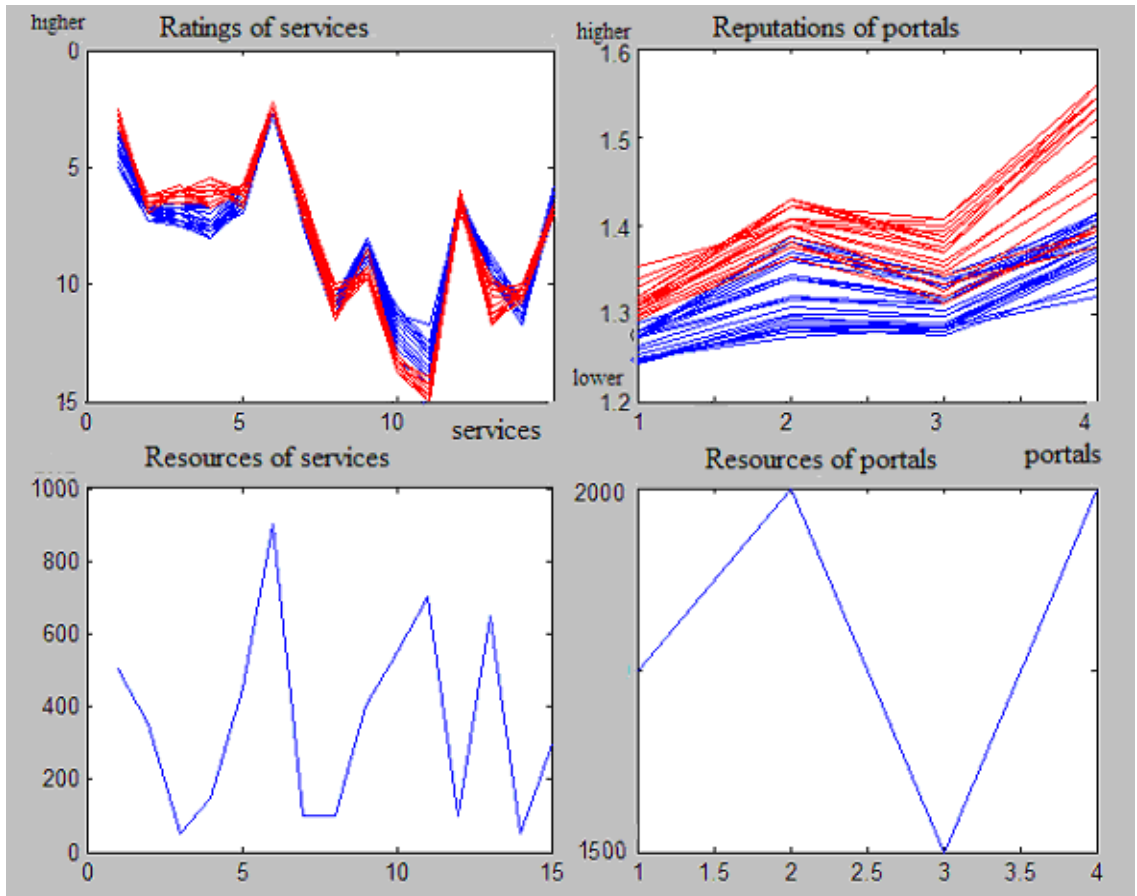

Figure 7: Distributions of services' ratings and resources, as well as portals' reputations and resources. Portal offer selection is random; the same strategy for services' acceptance by portals as above is used.

When the portals are randomly selected by the services, all portal reputations increase as a result of the process, which is unnatural (see Figure 7). With regards to services, we observe improved ratings for the initially highest rated services and worse ratings for the initially low rated services. Furthermore, the former, having a better rating, gets higher resultant resources and the latter, having lower rating, get lower resultant resources at the same time. Clearly, sacrificing resources to obtain better rating and vice versa, as well as sacrificing resources to preserve the reputation and vice verse, would follow our intuition, but this is not what we see at Figure 8.

It may seem unreasonable to a reader that if services choose the best portals they can benefit from, instead of choosing a portal randomly, then the initially best services will have their ratings depreciate while the initially worst services will gain in their ratings. However, the model plausibility should be achieved primarily on the level of individual agents: it is likely that an agent makes selections based on its own utility measure rather than performs planning for the whole multiagent community.

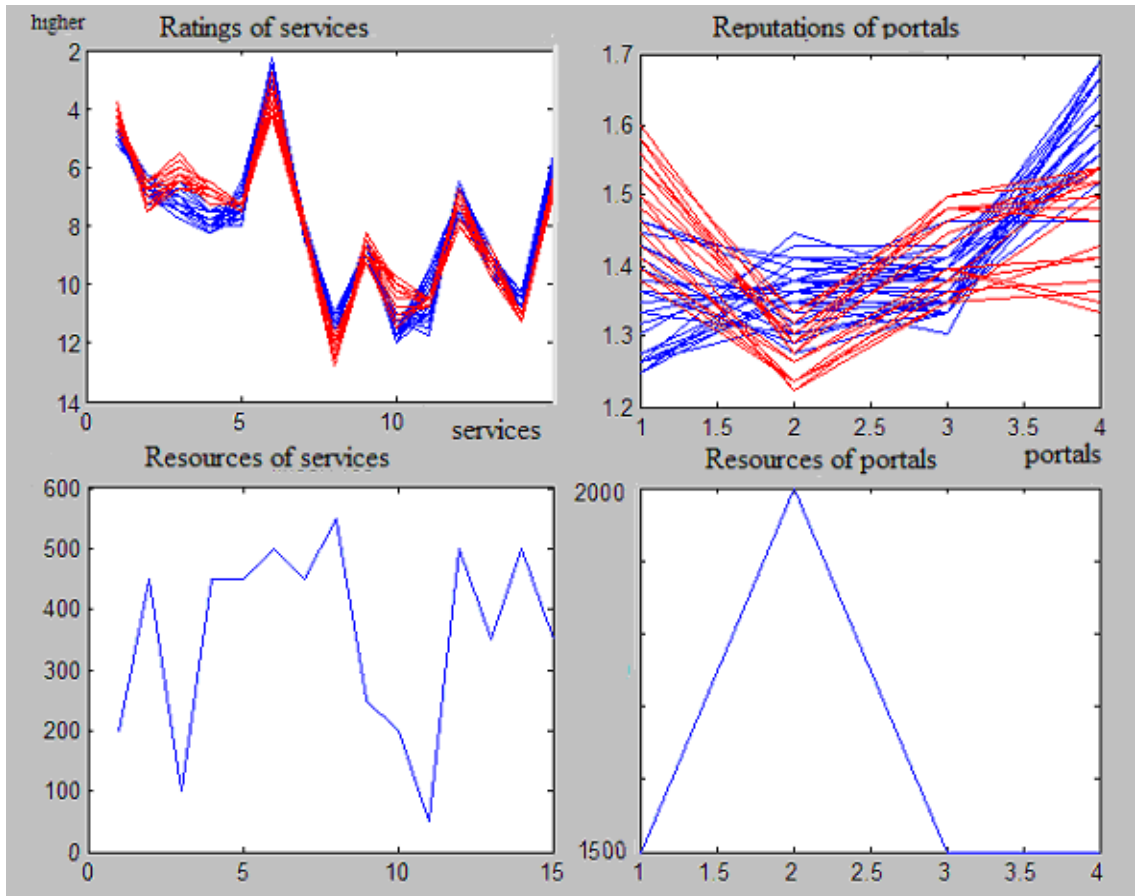

Figure 8: Distributions of ratings and resources. Both the portal offer selection and services' acceptance are random

In introducing the model of the subscription process, we are making assumptions which are as general as possible. These assumptions have resulted in observations regarding the sequence of characteristic zones in the evolution curves of ratings and resources of the involved agents. We therefore conjecture that an arbitrary subscription process that is connecting with rating-providing businesses would have a similar set of zones. Concerning the last zone, our conjecture is backed up by the assumption that this process eventually ends because the services would not want to spend any further resources. Since the rating, reputation and resources data for business agents is available, it is possible, in principle, to apply the respective feature extraction mechanism to identify the simulated process and its current zone.

## Results

In this study we have simulated the process of the interaction between the services which desire a higher rating on portals, whose revenue model is based on a subscription fee model where the flow of resources is from services to portals. We called this process the `subscription process'.

- When each agent participates in the subscription process, the reputation of independent portals, which do not accept subscriptions, drops. Also, the ratings of the highly-rated services, which choose not to subscribe to portals in order to compensate for subscriptions of other services, drop in the course of the process.

- When just a small portion of lowest-rated services offer subscriptions to portals, it nevertheless strongly decreases the reputation of portals accepting these subscriptions and the ratings of other services.

Therefore, it seems that when a low proportion of interacting agents participate in the subscription process, it has a negative effect on the ratings of others, and thereby encourages these other services to compensate for their lost rating by joining the process. At the same time, it is quite unprofitable with respect to both ratings and resources to stop subscribing to portals. For services, it would be profitable to stop subscribing synchronously, knowing that other services would cooperate and also stop subscribing. This is, however, impossible because the services do not have knowledge about each other in terms of participation in the subscription process.

In this study we suggested a possibility of how the natural intentions of services to sacrifice their resources to gain a batter rating may be formulated and the formulation of the intentions of portals to, possibly, sacrifice their reputation to gain resources from services, may compliment each other. We observed that the collective intentions of the above agents find the matching strategy, not the individual intentions of participating agents some of which may deviate from the majority of agents. In particular, initially highly rated services do not intend to enlist to the subscription process, but they have to accept the rules of the game once the other services have enrolled.

Since it is possible to observe real-world rating data and its evolution, one can extract the patterns of the subscription process, including the stationary zones and the transition zones. Such behaviour as oscillations in ratings , for example, will indicate that there is a strong competition between services for a particular portal. Such patterns can be revealed even analysing the search engine ranking resulting from keyword queries, although its mechanism and underlying processes are totally different[ML7].

## Discussion and Related Work

We have presented the process of competitive services *officially* subscribing to a rating mechanism on portals. In reality, this process may not have such a formal arrangement and occur in way where different participating agents lack information about the subscription arrangements of others. We have obtained the sequence of zones in our simulation process: transition from the initial zone to the final zone is expected to be associated with some *legalisation* process, when explicit rules of subscription offer/acceptance are formed and every agent becomes knowledgeable of these rules. The services subscription model should become transparent to the customers, and we suppose that some legislation will control the practice of this process and enforce the disclosure of its details. Currently, the Federal Trading Committee in USA recommends search engines having paid-placement advertising results to clearly separate these from results obtained from the search engine ranking algorithm (FTC 2004, www.ftc.gov/bcp/conline/pubs/buspubs/dotcom/).

We also think that the applicability of the above simulations goes beyond the online media. When the practice of subscription to the online rating services becomes generally accepted without clear alternatives, TV and paper media may wish to follow it.

Acceptability of the concept of monetary value associated with rating is not as striking as may seem to the reader initially. Consulting various media, the majority of people have got used to the idea that all the information is biased and therefore needs some re-digestion to be trustworthy. We believe that a rating portal, which prefers to stay independent, would not impress the audience as being so because the ratings of such a portal may significantly deviate from those of other portals involved in the subscription process.

The other possibility that needs to be mentioned is that the subscription process may become illegal. In this case the process of subscription laid out in this study may be perceived by the reader as a process of *corruption*. Since our simulations suggest the criteria to extract the behavioural patterns from the rating data that is publicly available, in this case, we may be able to reveal corruption that is specific to rating portals.

This study highlights the role of the concept of *distributed mental attitudes* for simulating the processes in a society. The concept of distributed knowledge have been thoroughly explored in the artificial intelligence literature and applied to a variety of multi-agent models (see Fagin et al 1995, Galitsky 2002). At the same time the notion of *distributed intentions* has not been extensively applied to the simulation of economical or social processes. In this study we may define distributed intentions as the intentions of the majority of community members that participate in a process such that other members are forced to participate as well even if they do not have direct explicit intentions of doing so. In other words, the collective intention of a multi-agent community to perform an action is the scenario where a majority of its (typical) members explicitly intend to perform the action, and the rest of (atypical) members are undecided whether or not to commit the action. If they do not perform the action then, believing that other agents will commit to it, the atypical agents will find their desired state (a long-term goal) further away from that of the typical agent.

The notion of distributed intention is worth applying to the setting of *multi-agent conflict* (Figure 9). In terms of a multi-agent conflict, the subscription process can be considered as a negotiation set out to achieve a state where the intentions of services becomes consistent with the intentions of portals. Note that the conflict of intentions between the services cannot be resolved. Without a subscription process there is no explicit conflict of intentions between the portals, but as only portals are competing for subscribing services, the conflict arises.

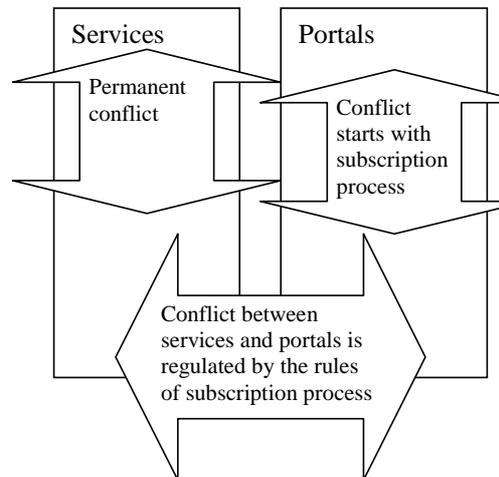

Figure 9: The outline of conflicts between the parties involved in subscription process.

[ML8] In this study we used numerical simulation to represent the subscription process, however the essence of our approach of obtaining the behavioural phenomenology should be referred to as *logical* instead. The simulation is concerned with the conflict resolution strategy, which is formed by participating agents in the online mode. Subscription process is a new form of economic behaviour

Coalition formation is a desirable behavior in a multiagent system, when a group of agents can perform a task more efficiently than any single agent can. Computational and communications complexity of traditional approaches to coalition formation, e.g., through negotiation, make them impractical for large systems. (Decker, Sycara and Williamson 1996) propose an alternative, physics-motivated mechanism for coalition formation that treats agents as randomly moving, locally interacting entities.

It is worth considering subscription process by a group of services as their coalition formation with rating portals. Coalition formation methods allow agents to join together and are thus necessary in cases where tasks can only be performed cooperatively by groups (Zacharia et al 1999, Lerman and Shehory 2000, Klusch and Gerber 2002) . This is the case in the Request For Proposal (RFP) domain, which is a general case for what we call here the subscription proposal. A requester business agent issues an RFP - a complex task comprised of sub-tasks - and several request processing agents need to join together to address this RFP. (Shehory and Kraus 1998) have developed a protocol that enables agents to negotiate and form coalitions, and provide them with simple heuristics for choosing coalition partners. The protocol and the heuristics allow the agents to form coalitions under the time constraints and incomplete information. The authors claim that the overall payoff of agents using suggested heuristics is very close to an experimentally measured optimal value, in accordance to their extensive experimental evaluation.

The results of our simulation study can be considered as creation of a novel advertising model that is suitable for online portals. Subscription process is a way of increasing demand by bringing the product to the attention of consumers. Advertising can be either informative or persuasive advertising. The effectiveness of advertising can be measured by the advertising elasticity of demand, which measures the percentage increase in demand divided by the percentage increase in advertising spending. In terms of advertisement, rating can be considered as a persuasive advertising means.

Large numbers of models have been used to assist in making advertising decisions. Econometric and other market models, as well as decision calculus models such as ADBUDG, have been used in the determination of advertising budgets (Little 1970 ). Media selection and scheduling models have included linear and nonlinear programming-type models, such as MEDIAC (Lodish 1966 ), and decision calculus models. Few models, such as ADMOD (Aaker 1977), have been designed to deal simultaneously with resources and media allocation decisions, where what we model in this study is a partial case of the latter. Our model can be considered as a complementary to the advertising model of (Vidale and Wolfe 1957). It is an econometric model that represents the rate of change of sales as a function of the rate of advertising spending. The tagged effect of advertising is incorporated using a sales decay term (The model allows the effect of advertising to have different rise versus decay rates).

Returning to the real-life problems, we cannot reject the possibility that the rating portals would form their business model in accordance to what we suggest in this paper. The question remains, if not the suggested business model, what else should the rating portals do nowadays to have a stable revenue stream?